\documentclass[conference,letterpaper]{IEEEtran}

\usepackage[utf8]{inputenc}
\usepackage[T1]{fontenc}
\usepackage{url}
\usepackage{ifthen}
\usepackage{cite}
\usepackage{amsfonts}
\usepackage{amssymb}
\usepackage{graphicx}
\usepackage{amsmath}
\usepackage{algorithmic}
\usepackage{algorithm}
\usepackage{array}
\usepackage[caption=false,font=normalsize,labelfont=sf,textfont=sf]{subfig}
\usepackage{textcomp}
\usepackage{stfloats}
\usepackage{url}
\usepackage{verbatim}
\usepackage{tabularx}
\usepackage{arydshln}
\usepackage{makecell}
\usepackage{fancyhdr}

\begin{document}
\title{Semantic Huffman Coding using Synonymous Mapping}

\author{%
    \IEEEauthorblockN
    {
        Jin~Xu\IEEEauthorrefmark{1},
        Kai~Niu\IEEEauthorrefmark{2},
        Zijian~Liang \IEEEauthorrefmark{1},
        and Ping~Zhang\IEEEauthorrefmark{2}
    }
    \IEEEauthorblockA
    {\IEEEauthorrefmark{1}%
        The Key Laboratory of Universal Wireless Communications, Ministry of Education
    }
    \IEEEauthorblockA
    {\IEEEauthorrefmark{2}%
        The State Key Laboratory of Networking and Switching Technology
    }
    \IEEEauthorblockA
    {
        Beijing University of Posts and Telecommunications, Beijing 100876, China \\
        Email: \{xujinbupt, niukai, liang1060279345, pzhang\}@bupt.edu.cn
    }
}

\maketitle
\thispagestyle{fancy}


\begin{abstract}
Semantic communication stands out as a highly promising avenue for future developments in communications. Theoretically, source compression coding based on semantics can achieve lower rates than Shannon entropy. This paper introduces a semantic Huffman coding built upon semantic information theory. By incorporating synonymous mapping and synonymous sets, semantic Huffman coding can achieve shorter average code lengths. Furthermore, we demonstrate that semantic Huffman coding theoretically have the capability to approximate semantic entropy. Experimental results indicate that, under the condition of semantic lossless, semantic Huffman coding exhibits clear advantages in compression efficiency over classical Huffman coding.
\end{abstract}

\section{Introduction}

Ever since C. E. Shannon introduced classical information theory (CIT) \cite{1948Shannon}, it has effectively steered the trajectory of modern communication technology for over seven decades. This theory establishes a comprehensive and internally coherent mathematical model of communication, grounded in probability theory. This enables the quantification of information, providing a systematic framework for analysis. Guided by CIT, numerous advanced coding and transmission technologies have been implemented. Nevertheless, with modern communication technology nearing the theoretical limit of CIT, the current research focus is on discovering novel communication theories capable of surpassing these limits.

In \cite{weaver1953recent}, Weaver delineated three levels of communication: the technical level, the semantics level, and the effectiveness level. Classical information theory only reaches the technical level. To enhance the performance of communication systems, it is imperative to integrate semantics into information theory. Fortunately, a large amount of work on semantic information has made great progress. Carnap \cite{carnap1952outline} considered using logical probability functions to describe and measure semantic information. Subsequently, Bao et al. \cite{bao2011towards} proposed a framework to measure the semantic information in information sources and communication channels, and made corresponding generalizations of Shannon's source coding theorem and channel coding theorem. Moreover, De Luca et al. \cite{de1993definition} introduced the concept of fuzzy sets to measure the entropy value of semantic information. In recent years, the advancement of artificial intelligence technology has led to numerous studies leveraging neural networks to deal with semantic information \cite{xie2021deep}. Some studies \cite{farsad2018deep} have introduced joint source-channel coding schemes for semantic communications, exploring innovative approaches in the field. In addition, semantic communication methods in other modalities and scenarios have also been studied \cite{zhang2023deepma,wang2022wireless,dai2022nonlinear,bourtsoulatze2019deep,zhang2023model}.

Despite this, a unified semantic information theoretical framework is still lack to provide guidance for existing work on semantic information. Based on this, Niu et al. \cite{niu2024Mathematical} proposed a mathematical principle of semantic information. This theory establishes the connection between syntactic information and semantic information through synonymous mapping, and further answers the question of how semantic information should be represented through the concept of synonymous sets. Utilizing synonymous mapping and synonymous sets, the framework also provides definitions for fundamental concepts such as semantic entropy, semantic mutual information, and semantic channel capacity. From these foundations, the semantic source coding theorem and the semantic channel coding theorem are derived. Within this context, the semantic source coding theorem elucidates that semantic source coding can achieve a superior compression rate compared to traditional source coding, without introducing any semantic distortion. This signifies the considerable potential that semantic source coding holds in the realm of source compression.

Building on this foundation, the paper introduces an extended coding method for Huffman coding \cite{huffman1952method}, namely semantic Huffman coding.  This approach incorporates the concept of synonyms and integrates synonymous mapping and synonymous sets into the classical Huffman coding. By leveraging the synonymous relationships within diverse syntactic information, leaf nodes in the same synonymous set are merged during the construction of the Huffman code tree. Consequently, the semantic Huffman code, devised through this methodology, attains a shorter average code length compared to the classical Huffman code while preserving semantic integrity, which approximates the semantic entropy limit.

\section{Semantic Information and Coding Theory}

In this section, we first provide a concise overview of basic concepts in semantic information theory, including synonymous mapping, synonymous sets, and semantic entropy. Subsequently, the description of the semantic source coding theorem is presented, along with an elucidation of the compression limit inherent in semantic variable-length coding.

\begin{figure}[t]
 \centering{\includegraphics[scale=0.69]{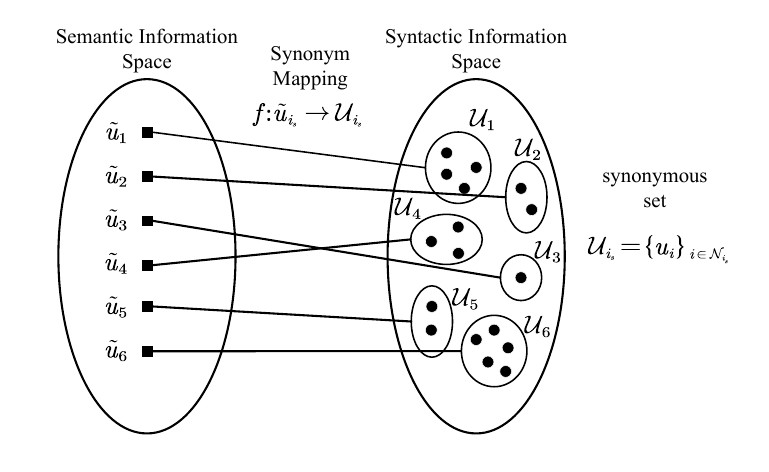}}
 \caption{Schematic diagram of the synonymous set and synonymous mapping.}\label{SynonymousSet}
\end{figure}

Assume that the syntactic source symbol space $\mathcal{U} = \left\{ {{u_1},{u_2}, \ldots ,{u_N}} \right\}$ and the semantic source symbol space $\tilde {\mathcal{U}} = \left\{ {{{\tilde u}_1},{{\tilde u}_2}, \ldots ,{{\tilde u}_{\tilde N}}} \right\}$ are connected through a synonymous mapping as shown in Fig \ref{SynonymousSet}. We define a set of source syntactic symbols sharing the same meaning as a synonymous set. Subsequently, synonymous mapping represents the process of mapping the syntactic symbols in the synonymous set into a semantic symbol, i.e., $f:{\tilde u_{{i_s}}} \to {\mathcal{U}}_{{i_s}}$. It should be noted that our focus is solely on the one-to-many mapping of semantic sets to syntactic sets. This also means that the size of $\tilde {\mathcal{U}}$ is no more than ${\mathcal{U}}$, i.e., $N \geqslant \tilde {N}$. The prior probability of a synonymous set should be the sum of the prior probabilities of its syntactic symbols,
\begin{equation}\label{probsum}
    p\left( {{\mathcal{U}_{{i_s}}}} \right) = \sum\limits_{i \in {\mathcal{N}_{{i_s}}}} {p\left( {{u_i}} \right)} ,
\end{equation}
where $\mathcal{N}_{{i_s}}$ represents the index set of syntactic symbols corresponding to the $i_s$-th synonymous set.

On the basis of synonymous mapping, the semantic entropy can be defined as
\begin{equation}
    {H_s}\left( {\tilde {\mathcal{U}}} \right) =  - \sum\limits_{{i_s} = 1}^{\tilde N} {p\left( {{\mathcal{U}_{{i_s}}}} \right)} \log p\left( {{\mathcal{U}_{{i_s}}}} \right) .
\end{equation}
Since the synonymous mapping is a one-to-many mapping, the semantic entropy is no more than the associated information entropy, that means ${H_s}\left( {\tilde {\mathcal{U}}} \right) \leqslant H\left( {\mathcal{U}} \right)$.

On this basis, the semantic source coding theorem can be derived using the asymptotic equipartition property. Define $R_s$ and $R$ as the rate of synonymous codeword set and the semantic code rate. For each code rate $R > {H_s}\left( {\tilde {\mathcal{U}}} \right)$ and , there exists a series of $\left( {{2^{n\left( {R + {R_s}} \right)}},n} \right)$ codes, when code length $n$ tends to sufficiently large, the error probability is close to zero. On the contrary, if $R < {H_s}\left( {\tilde {\mathcal{U}}} \right)$, then for any code, the error probability tends to 1 with $n$ sufficiently large.

Under the guidance of this theory, the semantic Kraft inequality can be obtained as following. For any prefix code over an alphabet of size $F$ exists if and only if the codeword length ${l_1},{l_2}, \ldots ,{l_{\tilde N}}$ satisfies
\begin{equation}
    \sum\limits_{{i_s} = 1}^{\tilde N} {{F^{ - {l_{{i_s}}}}} \leqslant 1} .
\end{equation}

Additionally, the semantic variable-length coding theorem provides that the optimal average code length of variable length codes ${{\bar L}^*}$ should satisfy
\begin{equation}\label{SemanticVarlen}
    \frac{{{H_s}\left( {\tilde {\mathcal{U}}} \right)}}{{\log F}} \leqslant {{\bar L}^*} < \frac{{{H_s}\left( {\tilde {\mathcal{U}}} \right)}}{{\log F}} + 1.
\end{equation}
This implies that, similar to the limit of classical variable-length coding being information entropy, the rate limit of semantic variable-length coding is the semantic entropy of the source.

\begin{figure}[t]
 \centering{\includegraphics[scale=0.52]{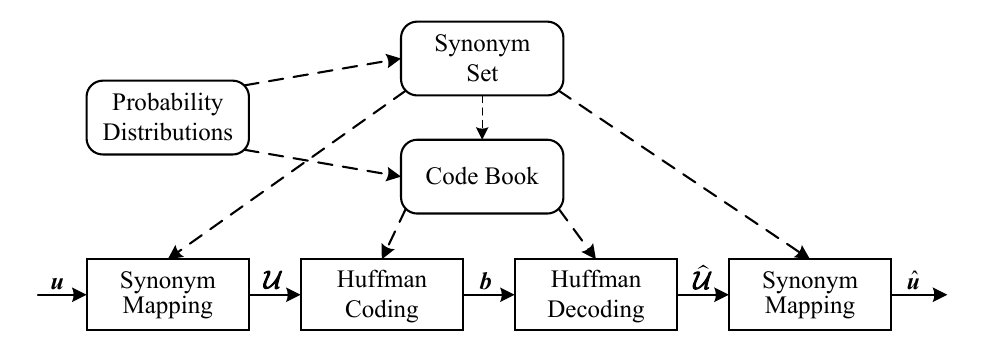}}
 \caption{The system block chart of the semantic Huffman coding.}\label{SystemBlock}
\end{figure}

\section{Semantic Huffman Coding}

In this section, we will elaborate on the principles and structure of semantic Huffman coding and make a preliminary exploration of its performance.

Similar to classic Huffman coding, semantic Huffman coding also comprises two main steps: (1) Constructing code trees and codebooks; (2) Encoding and decoding during coding procedure. The system block diagram of semantic Huffman coding is shown in the Fig \ref{SystemBlock}. The solid line represents coding procedure, and the dotted line represents support. Before coding procedure, the Huffman codebook needs to be generated based on the statistical characteristics of the source and the synonymous sets. By default, both the sending and receiving ends share the same codebook. During coding procedure, the sending end performs synonymous mapping and encoding after obtaining the source sequence to generate the sending codeword. The reverse operation is performed at the receiving end to restore the original information sequence through decoding and synonymous mapping.

\subsection{Semantic Huffman Code Tree}

The construction of code trees and codebooks is a key step for Huffman coding to achieve efficient information compression, providing a reliable and compact solution for transmission. In classical Huffman coding, the construction of the code tree is based on the statistical properties of the source. In semantic Huffman coding, we integrate the synonymous characteristics of the source into the construction process of the code tree through the creation of synonymous sets.

For the discrete source, first obtain the prior probability of each element through statistical analysis or modeling. Subsequently, synonymous mapping and synonymous set construction are performed based on the semantic characteristics of the source. According to the semantic characteristics of the source, syntactic symbols with the same or similar semantics are divided into the same set. In the process of dividing, synonymous sets and synonymous mapping should have the following characteristics: (1) Each synonymous set contains at least one syntactic symbol. (2) Different synonymous sets are disjoint with each other. (3) The union of all synonymous sets is equivalent to the set of syntactic symbols. These can be expressed as
\begin{equation}\label{SynonymousSetCondition}
    \left\{ {\begin{array}{*{20}{l}}
  {\forall {i_s}, {\mathcal{U}}_{{i_s}} \ne \oslash } \\
  {\forall {i_s} \ne {j_s}, {\mathcal{U}}_{{i_s}} \cap {{\mathcal{U}}_{{j_s}}} = \oslash } \\
  {\bigcup_{{i_s} = 1}^{{N_{{i_s}}}}{\mathcal{U}}_{{i_s}} = U}
\end{array}} \right..
\end{equation}
Depending on the type of source, the criteria for synonymous concepts may differ. For example, in textual sources, synonyms can be considered as a synonymous set; In image sources, pixels with similar colors can be grouped into the same synonymous set. After completing the division, we can obtain the synonymous mapping and synonymous sets space for this source. Then,the probability of each synonymous set can be calculated according to \eqref{probsum}.

\begin{figure*}[t]
\setlength{\abovecaptionskip}{0cm}
\setlength{\belowcaptionskip}{-0cm}
  \centering{\includegraphics[scale=0.68]{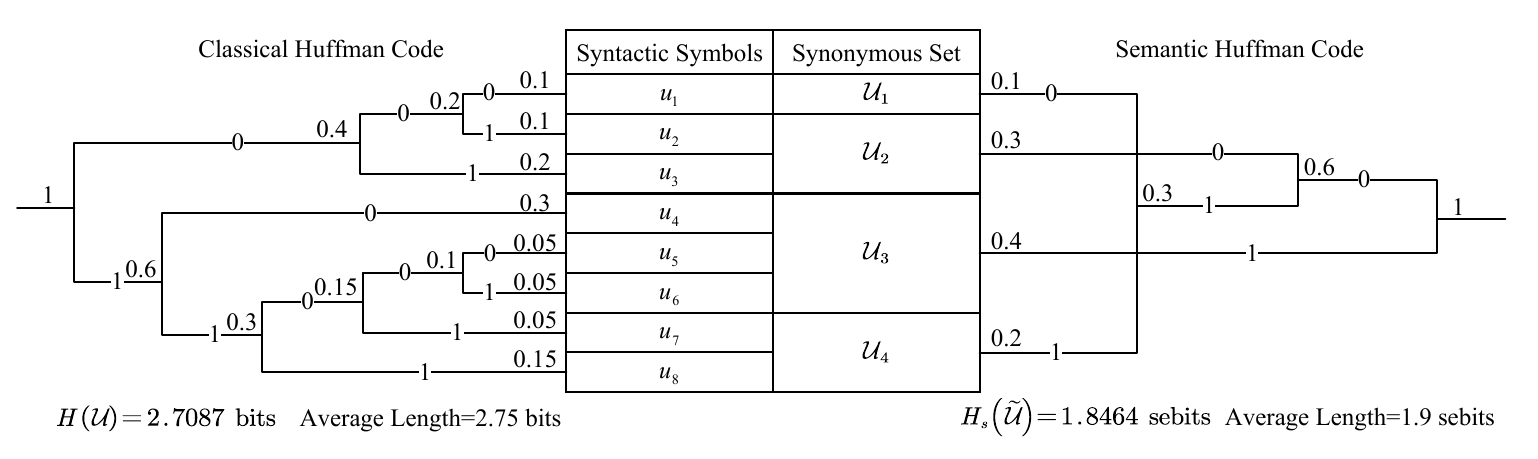}}
  \caption{Comparison of the average code length of semantic Huffman trees and syntactic Huffman trees, in which "sebits" denotes semantic bits for the resulting unit of semantic source coding, presented by \cite{niu2024Mathematical}.}\label{Tree}
  \vspace{-0em}
\end{figure*}

On this basis, we can further construct a semantic Huffman code tree. Different from classical Huffman coding, semantic Huffman code trees are constructed based on the prior probabilities of synonymous sets instead of the probabilities of syntactic symbols.
When constructing the code tree, first select the two synonymous sets ${\mathcal{U}_i}$ and ${\mathcal{U}_j}$ with the smallest prior probability as child nodes, and merge the two nodes into a parent node, recorded as ${\mathcal{U}_k}$. The prior probability of ${\mathcal{U}_k}$ is the sum of the probabilities of the two child nodes
\begin{equation}
    p\left( {{\mathcal{U}_k}} \right) = p\left( {{\mathcal{U}_i}} \right) + p\left( {{\mathcal{U}_j}} \right).
\end{equation}
Three nodes together form a branch. If there are ${\mathcal{U}_i}$ and ${\mathcal{U}_j}$ nodes in the existing code tree, combine the two nodes toward the root to form ${\mathcal{U}_k}$; otherwise, the corresponding new node is created and combined toward the root to form ${\mathcal{U}_k}$. Subsequently, the nodes corresponding to ${\mathcal{U}_i}$ and ${\mathcal{U}_j}$ are removed from the node set to be allocated, meanwhile the node corresponding to ${\mathcal{U}_k}$ is added to the node set. Repeat the above process until there is only one node left in the node set, completing the construction of the semantic Huffman code tree.

According to the Huffman code tree, starting from the root node, the Huffman codebook is constructed as follows. Designate the code "0" (or "1") for the upper branch and "1" (or "0")  for the lower branch. Each path from the root node to the leaf node corresponds to a symbol and the bits on the path are combined in order to construct the codeword. The correspondence between the synonymous set and the encoding codeword constitutes the semantic Huffman codebook.

Fig. \ref{Tree} illustrates an example between a classical Huffman code tree and a semantic Huffman code tree. The eight syntactic symbols ${u_1}, \ldots ,{u_8}$ are divided into four synonymous sets through synonymous mapping, i.e., ${\mathcal{U}_1} = \left\{ {{u_1}} \right\}$, ${\mathcal{U}_2} = \left\{ {{u_2},{u_3}} \right\}$, ${\mathcal{U}_3} = \left\{ {{u_4},{u_5},{u_6}} \right\}$ and ${\mathcal{U}_4} = \left\{ {{u_7},{u_8}} \right\}$.
The probability corresponding to each symbol and synonymous set is labeled on the both side of the table, in which the probability of each synonymous set equals to the sum of the probabilities of its syntactic elements. For example, $p\left( {{\mathcal{U}_2}} \right) = p\left( {{u_2}} \right) + p\left( {{u_3}} \right) = 0.1 + 0.2 = 0.3$. Building on this foundation, a classical Huffman code tree and a semantic Huffman code tree are constructed based on the probabilities associated with syntactic symbols and synonymous sets. The construction method follows the aforementioned process. Here we designate the code "0" for the upper branch and "1" for the lower branch. The final codes are formed by the arrangement of these branches along the path from the root node to the leaf nodes. For example, the codes of the syntactic symbol $u_2$ and the synonymous set $\mathcal{U}_2$ are $\left( 001 \right)$ and $\left( 00 \right)$, respectively. we can obtain the average code length of the classical Huffman Code and semantic Huffman code ${{\bar L}} = 2.75$ bits and ${{\bar L}_s} = 1.9$ sebits, respectively. For comparison, the corresponding information entropy  and semantic entropy are $H\left( \mathcal{U} \right) = 2.7087$ bits and ${H_s}\left( \tilde{ \mathcal{U}} \right) = 1.8464$ sebits.

This example indicates that semantic Huffman coding have advantages over classical Huffman coding in terms of the average code length. In fact, the semantic Huffman code tree can be seen as merging leaf nodes of the classical Huffman code tree based on synonymous sets. Therefore, it is foreseeable that it can achieve an advantage in average code length.

Herein, we prove the limit code length of semantic Huffman code corresponds to the semantic entropy ${H_s}\left( {\tilde {\mathcal{U}}} \right)$ of the source. Based on the similar induction method used for optimality proofs in classical Huffman coding \cite{cover1999elements}, we can simply prove the optimality of semantic Huffman coding by replacing symbolic probabilities $p \left( u_i \right)$ with synonymous set probabilities $p \left( \mathcal{U}_i \right)$. On this basis, the aforementioned semantic variable-length coding theorem can be used to demonstrate that the average code length of the semantic Huffman code conforms to \eqref{SemanticVarlen}, thereby proving that the semantic Huffman code can approach the semantic entropy limit.

\subsection{Semantic Huffman Coding and Decoding}

After completing the construction of the semantic Huffman code tree and codebook, Huffman encoding and decoding can be implemented during transmission.
Assuming that the length of the information sequence ${\boldsymbol{u}} = \left[ {{u_1},{u_2}, \ldots ,{u_M}} \right]$ sent by the source is $M$. According to the synonymous mapping, each symbol is mapped to its corresponding synonymous set, denoted as ${u_i} \in {\mathcal{U}_{{i}}}$. After completing the mapping, the synonymous set sequence $\boldsymbol{\mathcal{U}} = \left[ {{\mathcal{U}}_1,{\mathcal{U}}_2, \ldots ,{\mathcal{U}_M}} \right]$ can be obtained.

For each synonymous set ${\mathcal{U}_i}$ in the sequence $\boldsymbol{\mathcal{U}}$, its corresponding codeword can be obtained according to the codebook. During the encoding process, the corresponding codewords $\boldsymbol{b}_i$ are sequentially connected according to the order of the synonymous sets ${\mathcal{U}_i}$ to obtain the final encoding sequence $\boldsymbol{b}=\left(\boldsymbol{b}_1, \ldots, \boldsymbol{b}_M \right)$.

Since the Huffman code is a uniquely decodable code, each codeword is uniquely translatable. After obtaining the received sequence, the receiving end can decode $\boldsymbol{b}$ according to the code tree. The receiving end starts from the root node of the code tree and process decoding according to the bit instructions in $\boldsymbol{b}$. Each time a bit is read from $\boldsymbol{b}$, one step is taken toward the leaf node according to this bit. When reaching the leaf node, the decoding synonymous set $\hat {\mathcal{U}}_i$ corresponding to the node is output. Then return to the root node for the next decoding. The decoding sequence $\hat{\boldsymbol{\mathcal{U}}}$ can be obtained by arranging the synonymous sets output during the decoding process in order.

Since the synonymous set cannot be used as the final output, we still need objective syntactic symbols for expression. Therefore, it is necessary to select a syntactic symbol from each synonymous set ${\hat u}_i \in \hat {\mathcal{U}}_i$, and combine them into the final decoding sequence ${\hat {\boldsymbol{u}}}$. Obviously, in this process, the selected syntactic symbols ${\hat u}_i$ may not be the same as the original symbols ${u_i}$. This results in certain syntactic distortions. However, since  ${u_i}$ and ${{\hat u}_i}$ come from the same synonymous set, their semantics are the same. This implies that while decoding results may exhibit syntactic distortion, they remain semantically undistorted. In actual selection, in order to reduce the syntactic distortion, the syntactic symbol with the highest priori probability in the synonymous set can be selected as the output.

\section{Experiment Results}

In this section, we will validate the compression performance and semantic distortion-free characteristics of our proposed semantic Huffman coding.

Leveraging synonymous characteristics, we compress the text using semantic Huffman coding. The text example selects Shannon's famous paper ``\emph{A Mathematical Theory of Communication}''\cite{1948Shannon}. For ease of handling, we only encode the textual parts and filter out certain formulas and images. To ensure the semantic distortion-free characteristics of the text, we do not rely on existing synonym libraries. Instead, we manually construct a targeted thesaurus for synonymous mapping. The thesaurus satisfies all the requirements for synonymous mapping outlined in (\ref{SynonymousSetCondition}), while striving to ensure that words within synonymous sets share similar or closely related semantic meanings. In this process, we categorize both singular and plural forms of nouns into synonymous sets. Additionally, we consider words with different cases (uppercase and lowercase) as synonyms. Certain grammatical-specific words, like the ``\emph{be}'', ``\emph{am}'', ``\emph{is}'', ``\emph{are}'' which do not significantly impact semantics, are also treated as synonyms. Furthermore, words sharing similar meanings such as ``\emph{transmit}'' and ``\emph{send}'' are grouped into synonymous sets. A portion of the synonymous sets constructed for the example is given in Table \ref{1table}

\begin{table}[t]
  \centering
  \caption{Parts of synonymous sets used in the example.}\label{1table}
  \small %
  \renewcommand{\arraystretch}{1.2}
  \begin{tabular}{|c|c|}
    \hline
    Index & Synonymous sets \\
    \hline
    1 & be, am, is, are \\
    \hline
    2 & the, a, an \\
    \hline
    3 & send, transmit \\
    \hline
    4 & symbol, symbols \\
    \hline
    5 & message, information, messages\\
    \hline
    ... & ... \\
    \hline
  \end{tabular}
\end{table}

Building upon this foundation, we applied both classical Huffman coding and semantic Huffman coding to the article, and the outcomes are presented in Table \ref{2table}. The code length after applying semantic Huffman coding with synonymous mapping is reduced by 23,679 sebits compared to classic Huffman coding, accounting for approximately 6.0446\% of the total code length. It is worth noting that the average code lengths of classical Huffman coding and semantic Huffman coding approach the  information entropy and semantic entropy, respectively. This observation further validates the theoretical limits of both methods. For another viewpoint, it is evident that semantic Huffman coding has the capability to surpass the constraints imposed by information entropy. Additionally, the codebook of semantic Huffman coding is more streamlined. Assuming one letter occupies 8 bits and one binary code occupies 1 bit, then the semantic codebook and synonym set together occupy a total of 237,049 bits. Compared to the classical codebook with 257,712 bits occupation, there is an additional compression of approximately 8\%.

\begin{table}[t]
  \centering
  \small
  \caption{Compression results of classical Huffman coding and semantic Huffman coding.}\label{2table}
  \renewcommand{\arraystretch}{1.5}
  \begin{tabular}{|c|c|}
    \hline
    Item & Numerical results \\
    \hline
    \makecell{Classical Huffman code length \\ (Average length)} & \makecell{391735 bits\\(8.7762 bits)} \\
    \hline
    \makecell{Semantic Huffman code length \\ (Average length)} & \makecell{368056 sebits\\(8.2457 sebits)} \\
    \hline
    Information entropy & 8.7430 bits \\
    \hline
    Semantic entropy & 8.2119 sebits \\
    \hline
  \end{tabular}
\end{table}

Additionally, we also assess the semantic distortion of Huffman coding. We employ bilingual evaluation understudy (BLEU) \cite{papineni2002bleu} and word error rate (WER) \cite{farsad2018deep} metrics to characterize the syntactic distortion of the decoded text. Besides, bidirectional encoder representations from transformers (BERT) similarity \cite{kenton2019bert,xie2021deep} is utilized to reflect its semantic similarity with the original text. We conducted tests with two decoding schemes for comparison: one based on selecting according to maximum probability and the other involving random selection from the synonym set during the decoding stage. The results are presented in Table \ref{3table}, where the upper arrow indicates that a higher value is better, while the lower arrow indicates that a lower value is better.

\begin{table}[t]
  \centering
  \small
  \caption{Distortion measures for different decoding methods.}\label{3table}
  \renewcommand{\arraystretch}{1.5}
  \begin{tabular}{|c|c|c|}
    \hline
    Measure & \makecell{Maximum probability \\ selection} & Random selection \\
    \hline
    BLEU ($\uparrow$) & 54.6052 & 50.7667 \\
    \hline
    WER ($\downarrow$)& 0.2177 & 0.2688 \\
    \hline
    \makecell{BERT \\ similarity ($\uparrow$)} & 0.9969 & 0.8885 \\
    \hline
  \end{tabular}
\end{table}

From the table, it can be observed that while the text decoded by the semantic Huffman code does not exhibit significant scores in BLEU and WER, it attains a outstanding score in BERT similarity. This indicates that after synonymous mapping, there are noticeable differences in grammar between the decoded text and the original text, but it does not impact the overall semantics. This aligns with our proposed requirement for semantic distortion-free encoding. Additionally, symbol selection based on the maximum probability criterion can enhance overall coding performance and effectively mitigate ambiguity.

While BERT similarity does not attain a perfect score, this is primarily due to some limitations present in its semantic evaluation criteria. The expressions of the source channel coding theorem in both the original and translated texts are provided in \ref{4table} for clarification.

\begin{table}[t]
  \centering
  \caption{An example for the comparison of the original text and the reconstructed text.}\label{4table}
  \small %
  \renewcommand{\arraystretch}{1.75}
    \begin{tabular}{|m{1.8cm}|m{6cm}|}
    \hline
    \makecell{Original \\ text} & ``\emph{\textbf{Let a} source have entropy \textbf{H} (bits per symbol) and \textbf{a} channel have \textbf{a} capacity \textbf{C} (bits per second). \textbf{Then} it is possible to encode the output of the source in such \textbf{a way} as to \textbf{transmit} at \textbf{the} average rate \textbf{C/H} - $\epsilon$ \textbf{symbols} per second over the channel where $\epsilon$ \textbf{is} arbitrarily small. \textbf{It is} not possible to \textbf{transmit} at \textbf{an} average rate greater than C/H.}'' \\
    \hline
    \makecell{Reconstructed \\ text} & ``\emph{\textbf{let a} source have entropy \textbf{h} (bits per symbol) and \textbf{the} channel have \textbf{the} capacity \textbf{c} (bits per second). \textbf{next} it be possible to encode the output of the source in such \textbf{the methods} as to \textbf{send} at \textbf{a} average rate \textbf{c/h} - $\epsilon$ \textbf{symbol} per second over the channel where $\epsilon$ \textbf{are} arbitrarily small. \textbf{it be} not possible to \textbf{send} at \textbf{the} average rate greater than C/H.}'' \\
    \hline
   \end{tabular}%
\end{table}



In the example, bold font labeled in the sequences is employed to highlight the distinction between the two texts. Despite the numerous syntactic differences between the two passages, these differences do not impede our accurate understanding of their meaning, i.e., reliable transmission should be carried out under the condition that the source entropy is less than the channel capacity. Therefore, it can be asserted that the decoded text is semantic lossless compare to the original text.

\section{Conclusion}

Drawing upon semantic information theory, this paper introduces a groundbreaking semantic Huffman coding method capable of surpassing the Shannon limit. Diverging from classical Huffman coding, semantic Huffman coding combines leaf nodes during code tree construction through the use of synonymous mapping and synonymous sets. This innovative approach results in shorter average code lengths and heightened compression efficiency. Additionally, we demonstrate that the semantic Huffman coding can approach the theoretical semantic entropy limit. Through text-based simulations, we point out the advantages of semantic Huffman coding over classical Huffman coding in conditions free from semantic distortion.

\section*{Acknowledgement}

This work is supported by the National Natural Science Foundation of China (No. 62293481, 62071058).

\enlargethispage{-2.5cm}



\bibliographystyle{IEEEtran}
\bibliography{IEEEabrv,ref}

\end{document}